\documentclass[twoside,twocolumn,9pt]{article}
\usepackage{extsizes}
\usepackage[super,sort&compress,comma]{natbib}
\usepackage[version=3]{mhchem}
\usepackage[left=1.5cm, right=1.5cm, top=1.785cm, bottom=2.0cm]{geometry}
\usepackage{balance}
\usepackage{mathptmx}
\usepackage{sectsty}
\usepackage{graphicx} 
\usepackage{lastpage}
\usepackage[format=plain,justification=justified,singlelinecheck=false,font={stretch=1.125,small,sf},labelfont=bf,labelsep=space]{caption}
\usepackage{float}
\usepackage{fancyhdr}
\usepackage{fnpos}
\usepackage[english]{babel}
\addto{\captionsenglish}{%
  
}
\usepackage{array}
\usepackage{droidsans}
\usepackage{charter}
\usepackage[T1]{fontenc}
\usepackage[usenames,dvipsnames]{xcolor}
\usepackage{setspace}
\usepackage[compact]{titlesec}
\usepackage{hyperref}

\usepackage{comment}
\usepackage{booktabs}
\usepackage{multirow}
\usepackage{amsmath} 
\usepackage{wasysym}
\usepackage{xcolor}

\usepackage{epstopdf}

\definecolor{cream}{RGB}{222,217,201}
\definecolor{updatedpart}{RGB}{54,79,199}
\definecolor{notepart}{RGB}{224,49,49}

\begin{document}

\pagestyle{fancy}
\thispagestyle{plain}
\fancypagestyle{plain}{
\renewcommand{\headrulewidth}{0pt}
}

\makeFNbottom
\makeatletter
\renewcommand\LARGE{\@setfontsize\LARGE{15pt}{17}}
\renewcommand\Large{\@setfontsize\Large{12pt}{14}}
\renewcommand\large{\@setfontsize\large{10pt}{12}}
\renewcommand\footnotesize{\@setfontsize\footnotesize{7pt}{10}}
\makeatother

\renewcommand{\thefootnote}{\fnsymbol{footnote}}
\renewcommand\footnoterule{\vspace*{1pt}%
\color{cream}\hrule width 3.5in height 0.4pt \color{black}\vspace*{5pt}} 
\setcounter{secnumdepth}{5}

\makeatletter 
\renewcommand\@biblabel[1]{#1}            
\renewcommand\@makefntext[1]%
{\noindent\makebox[0pt][r]{\@thefnmark\,}#1}
\makeatother 
\renewcommand{\figurename}{\small{Fig.}~}
\sectionfont{\sffamily\Large}
\subsectionfont{\normalsize}
\subsubsectionfont{\bf}
\setstretch{1.125} 
\setlength{\skip\footins}{0.8cm}
\setlength{\footnotesep}{0.25cm}
\setlength{\jot}{10pt}
\titlespacing*{\section}{0pt}{4pt}{4pt}
\titlespacing*{\subsection}{0pt}{15pt}{1pt}

\fancyfoot{}
\fancyfoot[LO,RE]{\vspace{-7.1pt}\includegraphics[height=9pt]{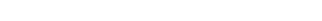}}
\fancyfoot[CO]{\vspace{-7.1pt}\hspace{13.2cm}\includegraphics{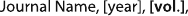}}
\fancyfoot[CE]{\vspace{-7.2pt}\hspace{-14.2cm}\includegraphics{head_foot/RF}}
\fancyfoot[RO]{\footnotesize{\sffamily{1--\pageref{LastPage} ~\textbar  \hspace{2pt}\thepage}}}
\fancyfoot[LE]{\footnotesize{\sffamily{\thepage~\textbar\hspace{3.45cm} 1--\pageref{LastPage}}}}
\fancyhead{}
\renewcommand{\headrulewidth}{0pt} 
\renewcommand{\footrulewidth}{0pt}
\setlength{\arrayrulewidth}{1pt}
\setlength{\columnsep}{6.5mm}
\setlength\bibsep{1pt}

\makeatletter 
\newlength{\figrulesep} 
\setlength{\figrulesep}{0.5\textfloatsep} 

\newcommand{\topfigrule}{\vspace*{-1pt}%
\noindent{\color{cream}\rule[-\figrulesep]{\columnwidth}{1.5pt}} }

\newcommand{\botfigrule}{\vspace*{-2pt}%
\noindent{\color{cream}\rule[\figrulesep]{\columnwidth}{1.5pt}} }

\newcommand{\dblfigrule}{\vspace*{-1pt}%
\noindent{\color{cream}\rule[-\figrulesep]{\textwidth}{1.5pt}} }

\makeatother

\twocolumn[
  \begin{@twocolumnfalse}
{\includegraphics[height=30pt]{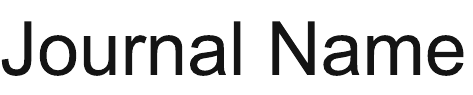}\hfill\raisebox{0pt}[0pt][0pt]{\includegraphics[height=55pt]{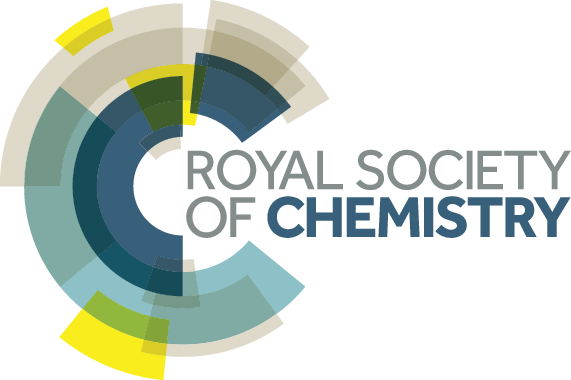}}\\[1ex]
\includegraphics[width=18.5cm]{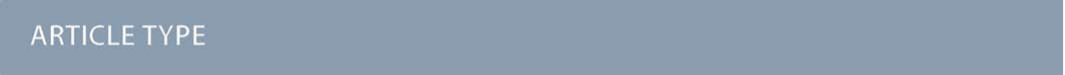}}\par
\vspace{1em}
\sffamily
\begin{tabular}{m{4.5cm} p{13.5cm} }

\includegraphics{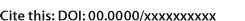} & \noindent\LARGE{\textbf{PIGNet2: A Versatile Deep Learning-based Protein--Ligand Interaction Prediction Model for Binding Affinity Scoring and Virtual Screening$^\dag$}} \\
\vspace{0.3cm} & \vspace{0.3cm} \\

 & \noindent\large{Seokhyun Moon,\textit{$^{a}$} Sang-Yeon Hwang,\textit{$^{b}$} Jaechang Lim,\textit{$^{b}$} and Woo Youn Kim$^{\ast}$\textit{$^{abc}$}} \\

\includegraphics{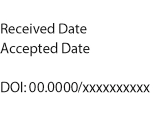} & \noindent\normalsize{Prediction of protein--ligand interactions (PLI) plays a crucial role in drug discovery as it guides the identification and optimization of molecules that effectively bind to target proteins. Despite remarkable advances in deep learning-based PLI prediction, the development of a versatile model capable of accurately scoring binding affinity and conducting efficient virtual screening remains a challenge. The main obstacle in achieving this lies in the scarcity of experimental structure--affinity data, which limits the generalization ability of existing models. Here, we propose a viable solution to address this challenge by introducing a novel data augmentation strategy combined with a physics-informed graph neural network. The model showed significant improvements in both scoring and screening, outperforming task-specific deep learning models in various tests including derivative benchmarks, and notably achieving results comparable to the state-of-the-art performance based on distance likelihood learning. This demonstrates the potential of this approach to drug discovery.
}

\end{tabular}

 \end{@twocolumnfalse} \vspace{0.6cm}

  ]

\renewcommand*\rmdefault{bch}\normalfont\upshape
\rmfamily
\section*{}
\vspace{-1cm}


\makeatletter\def\Hy@Warning#1{}\makeatother\footnotetext{\textit{$^{a}$~Department of Chemistry, KAIST, 291 Daehak-ro, Yuseong-gu, Daejeon, 34141, Republic of Korea. E-mail: wooyoun@kaist.ac.kr}}
\makeatletter\def\Hy@Warning#1{}\makeatother\footnotetext{\textit{$^{b}$~HITS Incorporation, 124 Teheran-ro, Gangnam-gu, Seoul, 06234, Republic of Korea.}}
\makeatletter\def\Hy@Warning#1{}\makeatother\footnotetext{\textit{$^{c}$~AI Institute, KAIST, 291 Daehak-ro, Yuseong-gu, Daejeon, 34141, Republic of Korea.}}

\footnotetext{\dag~Electronic Supplementary Information (ESI) available: [details of any supplementary information available should be included here]. See DOI: 00.0000/00000000.}


\section{Introduction}
Predicting protein--ligand interaction (PLI) plays a critical role in the early stages of drug discovery.\cite{PLI1,PLI2,PLI3}
It can be mainly utilized for two purposes: virtual screening to efficiently identify hit candidates from a large chemical space for a target protein and a process designed to refine these discovered molecules to increase their affinity for the target.
 The virtual screening emphasizes cost-effectiveness due to the extensive calculation required,\cite{LowCostSBVS1,LowCostSBVS2} while the binding affinity improvement process places greater emphasis on accuracy due to the need for precise evaluation of a relatively smaller number of molecules.\cite{LeadOpt1,LeadOpt2}
In this light, both fast and accurate PLI prediction is necessary to meet these requirements.
An ideal PLI prediction model should be computationally efficient and accurate in predicting binding affinity and thus be able to correlate the prediction with experimental binding affinities or correctly distinguish active molecules and decoy molecules.\cite{IdealSF1,IdealSF2}

Recently, deep learning-based models have emerged as a new avenue for PLI prediction.\cite{DeepDTA,DeepConvDTI,KDEEP,AK-Score}
Deep learning allowed for retaining fast computation speed while demonstrating better performance in predicting binding affinity for protein--ligand crystal structures.\cite{DLSF-GOOD1}
Despite their potential, most deep learning-based PLI prediction models are insufficient to be applied to various tasks at once.\cite{Challenge1,Challenge2,Challenge3,Challenge4}
Instead, they are task-specific, focusing only on scoring,\cite{AK-Score,OnionNet-2,Sfcnn} pose optimization,\cite{DeepDock,RTMScore,DeepRMSD} or screening.\cite{OnionNet-SFCT}
Specifically, the scoring task is to predict the binding affinity of protein--ligand complexes, and the screening task is to classify different compounds into true binders and non-binders.
Contrary to the common expectation that a model with high accuracy in binding affinity scoring will also have high accuracy in virtual screening, the performance of these two tasks is often at odds because deep learning models tend to learn exclusive (rather than generalizable) features to perform best at each task.
For example, models for predicting binding affinity trained only on crystal structures performed well at scoring crystal or near-native structures but struggled with tasks such as identifying specific binders to a target protein among diverse molecules or evaluating computer-generated structures as required in virtual screening.\cite{TaskSpecificEx}
Meanwhile, models that employ a $\Delta$-learning strategy with computer-generated data\cite{OnionNet-SFCT,AEV} or target the binding pose optimization\cite{DeepDock,RTMScore} have shown improved performance in virtual screening but failed to rank the binding affinities of different protein--ligand complexes adequately.
These challenges underscore the difficulty in designing a versatile PLI prediction model that can effectively handle diverse tasks.

Designing a versatile deep learning-based PLI prediction model performing well on both scoring and screening is challenged by the low generalizability of the model, which is mainly due to the lack of well-curated structure--affinity data.\cite{DataDeficiency1,DataDeficiency2,DataDeficiency3}
This challenge persists despite the gradually increasing availability of binding structure data from experiments.\cite{LimitedWellCuratedData}
To overcome this hurdle, one can impose a generalizable inductive bias throughout various tasks into the model.
In the previous study, we have shown that incorporating the physics of non-covalent molecular interaction as an inductive bias improves the generalization of a deep learning-based PLI prediction model.\cite{PIGNet}
In addition, data augmentation strategies can mitigate the problem of the lack of experimental structure--affinity data.
Previous approaches adopted data augmentation strategies by training the model with docking-generated structures to predict the binding affinity of non-binding structures to be lesser than that of experimental structures.\cite{CrossDocked2020,CrossDockImpact}
These strategies are based on the physical intuition that highly divergent structures of cognate ligands or structures with non-cognate ligands for a given target would have weaker binding affinities than a crystal structure.
However, models trained only with such augmented data have exhibited a relative decrease in scoring performance, diminishing their utility for this particular task.\cite{PIGNet,NDABAD}

Recently, GenScore\cite{GenScore} reported state-of-the-art performance on various tasks including scoring and screening.\footnote{Shortly before our submission, we became aware of the recently published GenScore. To ensure that our paper is comprehensive and up-to-date, we included a comparative analysis of our model with the results presented in the GenScore study.}
It is noteworthy that GenScore used neither physics-based inductive bias nor data augmentation, and simply focuses on learning the distance likelihood of binding structures instead of predicting their binding affinity.
Direct prediction of binding affinities has the great advantage that the predicted values can be directly compared to the experimental results, allowing for an intuitive explanation and directions for further improvement, whereas distance likelihood-based results only allow for relative comparisons between predicted values.

Here, we propose a versatile deep learning-based PLI prediction model by improving its generalization ability with physics-based inductive bias and data augmentation strategy.
Along with the previous data augmentation strategies, we generated near-native structures that are energetically and geometrically similar to crystal structures to consider their limited resolution and dynamic nature.
The model was then trained to predict the binding affinity of these structures to be the same as the experimental value, effectively recognizing that these structures lie on the same local minima as crystal structures.
As a result, PIGNet2, which is based on the physics-informed graph neural network modified from our previous work,\cite{PIGNet} showed significantly enhanced scoring and screening performance.

To demonstrate the potential applicability of PIGNet2 in scoring and screening tasks, we evaluated it against various benchmarks.
We used the CASF-2016 benchmark\cite{CASF-2016} to compare the overall performance on scoring, docking, and screening.
Following that, we also used DUD-E\cite{DUD-E} and DEKOIS2.0\cite{DEKOIS2.0} as widely-adopted virtual screening benchmarks to evaluate the screening performance in more detail.
As a further assessment of the scoring performance, we adopted two separate derivative benchmarks provided by \citet{DerivativeBenchmark} and the Merck FEP benchmark,\cite{merck} which will be called the derivative benchmark 2015 and 2020, respectively.
Both are the structure--affinity datasets of derivative compounds with various target proteins and are specifically designed for assessing the scoring performance of PLI prediction models between structurally similar molecules.
Overall, PIGNet2 outperformed task-specific models in all benchmarks, achieving results on par with the state-of-the-art performance of GenScore, while leaving room for further improvement thanks to its use of intuitively explainable physics. 
Thus, our approach provides an alternative solution to develop a versatile deep learning model that can be used for hit identification and lead optimization in drug discovery.

\section{Methods}
\label{sec:2}

\subsection{PDBbind dataset}
\label{sec:2.1}

The PDBbind dataset,\cite{PDBbind} which comprises protein--ligand binding complex data curated from the Protein Data Bank (PDB),\cite{RCSBPDB} is divided into general, refined, and core sets based on the strictness of the curation.
The core set is subjected to the most rigorous curation criteria, thus including representative entities with respect to the target protein.
A growing trend among recent PLI prediction models is to exploit the general set in order to leverage a larger pool of crystal structures for training.\cite{CrossDocked2020,CrossDockImpact}
This approach was inspired by a previous work demonstrating that using the larger general set can improve the performance of PLI prediction models when compared to the refined set. \cite{LowQualityData}

In our study, however, we employed the refined set to carry out data augmentation for all the proteins and ligands with limited computational resources.
This may indicate that our model still has room for improvement by expanding the number of crystal structure data with the general set.
Of the 5,312 complexes present in the refined set, we omitted the core set included in the CASF-2016 benchmark\cite{CASF-2016} and redundant complexes.
As a result, we ended up with a training set of 5,046 complexes and a test set of 266 complexes.

\subsection{Data augmentation strategies}
\label{sec:2.2}

In this section, we present our novel positive data augmentation (PDA) strategy in conjunction with various negative data augmentation (NDA) strategies: re-docking, random-docking, and cross-docking data augmentation.

\subsubsection{Positive data augmentation}
\label{sec:2.2.1}

Recent deep learning-based PLI prediction models commonly utilize crystal structures and docked structures from NDA as input, where the unnatural NDA-generated structures get substantially more abundant than crystal structures.
We hypothesize that this data imbalance potentially impedes the discrimination between near-native and unnatural (or unstable) structures, thereby degrading the performance of binding affinity prediction for crystal or near-native structures.
Since each crystal structure is a single snapshot of conformations around local minima of a potential energy surface, treating near-native structures as true binders may implicitly integrate the conformational ensemble effect into the PLI prediction model.
In this light, we introduced a novel data augmentation strategy, PDA, designed to generate energetically and geometrically near-native conformations for any given complex structure.

For PDA, we first generated 1,000 conformations of each ligand using the ETKDG conformer generation method.\cite{ETKDG}
We optimized those structures using the universal force field (UFF)\cite{UFF} and Merck molecular force field (MMFF).\cite{MMFF}
This can yield a maximum of 3,000 data points for each complex.
Next, the resulting structures were aligned to the ligand's pose in the crystal.
Finally, the structures are minimized using the Smina docking software,\cite{Smina} a forked version of AutoDock Vina,\cite{AutodockVina} to avoid clashes.
We then selected structures that satisfy two criteria: (1) a ligand root mean square deviation (RMSD) less than 2~\AA\ compared to the crystal structure and (2) a mean absolute error less than 1 kcal/mol between the Smina scores of the crystal structure and the generated structures.
The latter criterion aims to select structures energetically similar to the near-native structure, in addition to the former, the geometric criterion that is more generally used.
While the scoring function of Smina only approximates the PLI potential energy surface (PES), it is rational to regard structures with similar scores in a confined range as energetically near-native on the actual PES considering the continuity of energy.
Finally, to remove highly similar structures that can be considered duplicates, we additionally pruned the generated structures so that the RMSD between every pair of the generated structures is greater than 0.25 \AA.
Along with the above, structures generated by re-docking crystal structures using Smina were also used.
The RMSD between every pair of generated structures for each complex was maintained below 2 \AA.

\subsubsection{Negative data augmentation}
\label{sec:2.2.2}
The negative data augmentation strategies mostly follow the methods outlined in our previous work,\cite{PIGNet} with additional guidelines to generate non-binding structures more rigorously.
First, the re-docking data augmentation generates structures by docking ligands into a cognate target and then extracting unstable structures.
Based on the fact that crystal structures are stable binding poses, one can infer that ligand structures that deviate greatly from the crystal structure will be highly unstable.
Thus, we used docking-generated structures with the ligand RMSD greater than 4~\AA\ compared to the corresponding crystal structure.
Second, cross-docking data augmentation uses the idea that a non-cognate protein--ligand pair is less likely to form a bound complex.
To implement this, we grouped proteins based on a protein sequence similarity of 0.4 using the cd-hit software.\cite{cd-hit1,cd-hit2}
Then, pairs of different protein clusters were sampled, and for each pair of clusters, proteins from one cluster were docked with ligands from the other cluster to generate structures of non-cognate protein--ligand pairs.
Lastly, the random-docking data augmentation strategy assumes that an arbitrarily chosen molecule is unlikely to be a true binder to a given protein by chance.
We generated the corresponding structures by docking a random molecule from the IBS molecule dataset\cite{STOCKS} to each protein.
For all the negative data augmentation strategies, we used Smina for docking and structure minimization and the DockRMSD\cite{DockRMSD} software for calculating the ligand RMSD.

\subsection{Data preprocessing}
\label{sec:2.3}
To convert crystal structures and the simulated structures into inputs for the PLI prediction model, we carried out additional preprocessing steps.
We protonated all the protein structures with the Reduce software.\cite{reduce}
For the ligands, we protonated them at pH 7.4 using Dimorphite-DL.\cite{dimorphiteDL}
Water and hydrogens were removed from the complexes.
As the final step, only the protein residues containing heavy atoms within 5~\AA\ or less from the ligand were extracted and used as the protein pocket.
We used RDKit,\cite{RDKit} Open Babel,\cite{openbabel} and PyMOL\cite{PyMOL} throughout.
A detailed breakdown of the training and test sets, derived through each data augmentation, is presented in Table \ref{tab:data_counts}.

\begin{table}[h!]
\caption{
Number of data points generated from each data augmentation strategy
}
\centering
\begin{tabular}{lcc}
\toprule
Data augmentation strategy & Training set & Test set \\
\midrule
PDA & 375,184 & 21,377 \\
NDA (re-docking) & 254,163 & 12,109 \\
NDA (cross-docking) & 503,073 & 26,470 \\
NDA (random-docking) & 957,775 & 50,496 \\
\bottomrule
\end{tabular}
\label{tab:data_counts}
\end{table}

\subsection{Model architecture}
\label{sec:2.4}

PIGNet2 adopted the model architecture and physics terms from our previous work\cite{PIGNet} except for the initial atom features.
Refer to ESI$\dagger$ for the modified initial atom features of PIGNet2.

PIGNet2 works as follows: preprocessed pocket and ligand structures, in conjunction with input features, are first passed through a feedforward network and then through a gated graph attention network.
The resulting pocket and ligand features are then passed through an interaction network, which allows the embedding of additional information from the counterpart.
Following this, the pocket and ligand features are concatenated, facilitating the calculation of the contribution of physics terms to the binding affinity based on these unified features.

The total binding affinity $E^\mathrm{pred}$ predicted by the model is the sum of all intermolecular atom pairwise energies consisting of four terms: $E^\mathrm{vdW}$, $E^\mathrm{Hbond}$, $E^\mathrm{Metal}$, and $E^\mathrm{Hydrophobic}$.
Each of them represents intermolecular van der Waals (vdW), hydrogen bond, metal--ligand, and hydrophobic interactions, respectively. 
In order to incorporate the effect of entropy as regularization, the total energy is divided by $T^\mathrm{rot}$, a term proportionate to the number of rotatable bonds of the ligand.
The equation of the total energy is as follows:

\begin{equation}
E^\mathrm{pred}=\frac{E^\mathrm{vdW}+E^\mathrm{Hbond}+E^\mathrm{Metal}+E^\mathrm{Hydrophobic}}{T^\mathrm{rot}}.
\label{eq:1}
\end{equation}

One feature of our current model that differs from the previous is the inclusion of the Morse potential instead of the Lennard-Jones potential for $E_\mathrm{vdW}$.
The ability of PIGNet2 to score the binding affinity of crystal structures or to clearly distinguish between active and decoy molecules is highly dependent on the modeling of the vdW potential well.
For example, an overly broad potential well could result in the prediction of a degree of interaction even for atom pairs that are too far apart to contribute significantly to the interaction, resulting in predicting inherently unstable structures to be stable.
On the other hand, an excessively narrow potential well could lead to the prediction of repulsive vdW interactions for atom pairs that are appropriately close, resulting in predicting unstable energies for them.
The correct form of the potential well is, therefore, critical for accurate prediction.
Thus, directly adjusting the potential well offers significant advantages in the design and evaluation of deep learning-based physics-informed PLI prediction models.
However, the Lennard--Jones potential possesses insufficient flexibility to freely adjust the width of the potential well, which was our reason to introduce the derivative loss in the previous work.\cite{PIGNet}
As a more direct and precise alternative, we chose to use the Morse potential, which allows for explicitly controlling the potential well.
The formula for the Morse potential is given below:

\begin{equation}
E^\mathrm{vdW}=w\big((1-e^{-a(d-r)})^2-1\big),
\label{eq:2}
\end{equation}
where $d$ denotes the interatomic distance and $r$ is the sum of vdW radii of the atom pair.
Note that we used $r$ as the pairwise sum of vdW radii instead of the pairwise \textit{corrected} sum of vdW radii from the previous study\cite{PIGNet} not only for the Morse potential but also for all the other physics terms.
The coefficient $a$ modulates the width of the potential well, while $w$ affects the depth of the potential well.
For the case where $d$ is greater than $r$, the coefficient $a$ is predicted by the neural network, while for the opposite case, it is set as a hyperparameter, with a value of 2.1 chosen for PIGNet2.

\subsection{Training setup}
\label{sec:2.5}

\subsubsection{Loss function}
\label{sec:2.5.1}

PIGNet2 employs various loss functions to optimize the model through the learning objectives of each data augmentation strategy.
Specifically, the PDA and crystal structures, acting as near-native structures, are trained using the mean squared error loss.
This approach induces the model to precisely predict the experimental binding affinity for near-native structures as well as crystal structures to inform the model that both structures from PDA and crystal structures belong to the same local minima in PES.
In contrast, for data exhibiting significant structural deviations from the crystal structure, which is derived from the re-docking process of NDA, we employed a hinge loss to predict a lower binding affinity than that of the crystal structure.
Lastly, we applied a hinge loss for cross-docking and random-docking data augmentation in another way.
This helps to predict binding affinities for the structures higher than a criterion of -6.8 kcal/mol, consistent with the assumption that these structures are unlikely to have a binding interaction.
Altogether, the total loss function is a weighted sum of all the losses above.
A more detailed description of the loss function is shown in the ESI.$\dagger$

\subsubsection{Training procedure}
During the training with PDA, we merged the PDA and crystal structures into a single dataset.
Meanwhile, during the training with NDA, we set the number of data for each dataset that the model learns per epoch to be equal.
Throughout the training, we used a batch size of 64, a learning rate of 0.0004, and a dropout ratio of 0.1.
We used a single RTX A4000 GPU for all training and inference.
Finally, all our results are an ensemble of predictions from four models, each initialized with a different random seed.

\section{Results and discussion}
\label{sec:3}

\subsection{Performance on CASF-2016 benchmark}
\label{sec:3.1}

\begin{table*}[ht]
\caption{
Results on the CASF-2016 benchmark.
$\mathrm{EF}_{1\%}$, $\mathrm{SR}_{1\%}$, $R$, and $\rho$ are top 1\% enrichment factor, success rate, Pearson correlation coefficient, and Spearman rank correlation coefficient, respectively.
The results of all baseline PLI prediction models are originated from their respective literature.
For PIGNet2, we report the results of 4 randomly initialized model ensemble trained with both positive data augmentation and negative data augmentation
}
\resizebox{\textwidth}{!}
{
\begin{tabular}{lccccccc}
\toprule
\multirow{2}{*}{Model} &
  \multirow{2}{*}{Prediction target} &
  \multicolumn{2}{c}{Screening power} &
  \multicolumn{1}{c}{Docking power} &
  \multicolumn{1}{c}{Scoring power} &
  \multicolumn{1}{c}{Ranking power} \\
 & & \multicolumn{1}{c}{Average $\mathrm{EF}_{1\%}$} &
 \multicolumn{1}{c}{$\mathrm{SR}_{1\%}$} &
  \multicolumn{1}{c}{$\mathrm{SR}_\mathrm{1}$} &
  \multicolumn{1}{c}{$R$} &
  \multicolumn{1}{c}{$\rho$} \\
\midrule
  DeepDock\cite{DeepDock} & Distance likelihood & 16.4 & 43.9\% & 89.1\% & 0.460 & 0.425 \\
  GenScore (\texttt{GT\_ft\_0.5})\cite{GenScore} & Distance likelihood & 28.2 & 71.4\% & 97.6\% & 0.773 & 0.659 \\
  GenScore (\texttt{GatedGCN\_ft\_1.0})\cite{GenScore} & Distance likelihood & 23.5 & 66.1\% & 95.4\% & 0.834 & 0.686 \\
\midrule
  OnionNet-SFCT(Vina)\cite{OnionNet-SFCT} & $\Delta$ binding affinity & 15.5 & - & 93.7\% & 0.428 & 0.393 \\
  $\Delta$-AEScore\cite{AEV} & $\Delta$ binding affinity & 6.16 & 19.3\% & 85.6\% & 0.740 & 0.590 \\
\midrule
  OnionNet-2\cite{OnionNet-2} & Exact binding affinity & - & - & - & 0.864 & - \\
  AEScore\cite{AEV} & Exact binding affinity & - & - & 35.8\% & 0.830 & 0.64 \\
  AK-score\cite{AK-Score} & Exact binding affinity & - & - & 36.0\% & 0.812 & 0.670 \\
  Sfcnn\cite{Sfcnn} & Exact binding affinity & - & - & 34.0\% & 0.795 & - \\
\midrule
PIGNet2 & Exact binding affinity & 24.9 & 66.7\% & 93.0\% & 0.747 & 0.651 \\
\bottomrule
\end{tabular}
}
\label{tab:CASF-2016}
\end{table*}

\subsubsection{CASF-2016 benchmark}
\label{sec:3.1.1}
To demonstrate the versatility of PIGNet2 for broad applications, we employed the well-established CASF-2016 benchmark.\cite{CASF-2016}

The CASF-2016 benchmark was carefully curated from 285 protein--ligand complexes in the PDBbind core set.
This benchmark provides a comprehensive set of four metrics: scoring power, ranking power, docking power, and screening power.
Each of these metrics serves a unique purpose in assessing PLI prediction models.

The metrics fall into two main categories.
The first category evaluates the ability of the model to predict binding affinity for crystal structures.
The second category evaluates the ability of the model to distinguish true-binding structures from a variety of computer-generated structures.
Scoring power and ranking power fall into the first category, and docking power and screening power fall into the second category.
The four powers together comprehensively evaluate the performance of models in different aspects of PLI prediction.

To be specific, the scoring power evaluates the ability of the model to predict the binding affinity of protein--ligand crystal structures and is assessed using the Pearson correlation coefficient $R$.
The ranking power measures the ability of the model to rank the binding affinities of protein--ligand complexes grouped by protein similarity.
It is evaluated using the Spearman rank correlation coefficient $\rho$.
The docking power assesses the ability of the model to identify near-native structures from computer-generated decoy structures.
The metric is evaluated based on a top $N$ success rate, $\mathrm{SR}_{N}$, where a case is considered successful if at least one of the top $N$ predicted structures for each complex has a ligand root mean square deviation (RMSD) of less than 2 \AA~when compared to the crystal structure.
Finally, the screening power evaluates the ability of the model to identify cognate protein--ligand complexes that can form a binding interaction among the vast amount of non-cognate protein--ligand complexes in cross-docking scenarios.
The screening power is assessed with the top $\alpha$ \% enrichment factor, $\mathrm{EF}_{\alpha \%}$, which is a measure of the ratio of active molecules included in the top $\alpha$ \% model predictions to the total number of active molecules, defined as follows:
\begin{equation}
\label{eq:enrichment_factor}
\mathrm{EF}_{\alpha\%}=
\frac{\mathrm{NTB}_\alpha}{\mathrm{NTB}_\mathrm{total}\times \alpha},
\end{equation}
where $\mathrm{NTB}_\alpha$ is a number of active molecules in top $\alpha$ \% and $\mathrm{NTB}_\mathrm{total}$ is total number of active molecules in overall dataset.
Along with $\mathrm{EF}_{\alpha \%}$, we also report the top $\alpha$ \% success rate, $\mathrm{SR}_{\alpha
 \%}$, which measures the success rate of finding the best binder among the top $\alpha$ \% top-ranked structures for all targets.

\subsubsection{Baseline models}
\label{sec:3.1.2}
We selected several task-specific deep learning-based PLI prediction models as baselines for comparative studies.
These models differ in their prediction targets during training and inference, tailored to excel in their respective objective tasks.
Thus, the models are categorized based on their prediction targets: distance likelihood, $\Delta$ binding affinity, and exact binding affinity, to differentiate the results from previous approaches better.
Our model, PIGNet2, falls into the category that directly predicts exact binding affinity.

DeepDock\cite{DeepDock} primarily aims to optimize protein--ligand structures.
Instead of predicting binding affinities, DeepDock predicts the distance likelihood of given structures by utilizing a mixture density network\cite{MDN} to model the statistical potential of protein--ligand structures.
GenScore,\cite{GenScore} a model built on the same formulation as DeepDock, was trained with an additional loss term to learn the correlation between the binding affinities of different complexes.
Since the authors provided ten models with different adjustable parameters and each showed different performances in various tasks, here we report the results of \texttt{GT\_ft\_0.5} and \texttt{GatedGCN\_ft\_1.0}, which respectively showed the best performances in screening and scoring.

Unlike previous, OnionNet-SFCT\cite{OnionNet-SFCT} and $\Delta$-AEScore\cite{AEV} estimate the final energy by a linear combination of correction terms to the Autodock Vina\cite{AutodockVina} scores.
These methods incorporate various computer-generated structures in their training process to enhance performance in virtual screening tasks.

Finally, baseline models that directly predict exact binding affinity include AK-score,\cite{AK-Score} Sfcnn,\cite{Sfcnn} OnionNet-2,\cite{OnionNet-2} and AEScore.\cite{AEV}
AK-score and Sfcnn use 3D convolutional neural networks (CNN), while OnionNet-2 uses a 2D feature map with a 2D CNN.
AEScore predicts binding affinity using a feedforward neural network based on the atomic environment vector representation.
By including these different models in our comparison, we can thoroughly evaluate the relative performance and robustness of PIGNet2.

\subsubsection{Performance of PIGNet2}
\label{sec:3.1.3}
The performance of PIGNet2 is shown in Table~\ref{tab:CASF-2016} along with all known results from the baseline models.
DeepDock, a model that predicts distance likelihood, showed excellent performance in distinguishing crystal structures from computer-generated structures, as evidenced by their high docking and screening powers.
However, its ability to predict or compare the binding affinity of crystal structures is limited.
It is natural because their distance likelihood computation can only infer the relative stability of conformations of a single protein--ligand complex.
This makes it difficult to compare different protein--ligand complexes of stable structures, mirroring the limitations of traditional knowledge-based PLI prediction models that utilized statistical potentials.
GenScore overcomes this difficulty by introducing an additional loss term that guides it to learn the correlation between experimental binding affinity and the statistical potential based on the distance likelihood, thereby achieving state-of-the-art performance in both scoring and screening tasks.

On the other hand, models that focus on accurate regression of binding affinities, which have been the subject of consistent research, generally demonstrate their strong performance in scoring and ranking.
Nevertheless, their performance on virtual screening-related metrics lags behind those predicting distance likelihoods.
Prime examples of such models include AEscore and OnionNet-2, and they have attempted to compensate for the poor screening performance by introducing various computer-generated structures and $\Delta$-learning.
However, although the resulting OnionNet-SFCT and $\Delta$-AEScore enhanced docking and screening power, their scoring and ranking powers were significantly reduced.
This trend was particularly pronounced for OnionNet-SFCT, indicating that designing versatile deep learning-based PLI prediction models is challenging even with data augmentation and $\Delta$-learning.

PIGNet2 aimed to perform equally well in all performance criteria by using a physics-informed graph neural network coupled with various data augmentation strategies.
Indeed, PIGNet2 showed high performance for all the metrics, comparable to the state-of-the-art performance of GenScore, while GenScore performed slightly better on each metric depending on the version.
It even outperformed the result of DeepDock in its primary objective, \textit{i.e.}, pose optimization, as shown by the docking power.
Specifically, PIGNet2 attained scoring and ranking powers comparable to OnionNet-2, AEScore, AK-score, and Sfcnn, all of which aim to score the binding affinity of crystal structures accurately.
PIGNet2 outperformed in all metrics, compared to OnionNet-SFCT and $\Delta$-AEScore, which aim to improve docking and screening performance by introducing the $\Delta$-learning strategy.
These results suggest that PIGNet2, trained with data augmentation strategies, can serve as a versatile deep learning-based PLI prediction model.

\subsection{Performance on classifying active and decoy compounds}
\label{sec:3.2}
While the CASF-2016 screening benchmark is well-designed, it slightly differs from real-world virtual screening scenarios due to relatively fewer numbers of actives and decoys.
In this light, we utilized the Directory of Useful Decoys-Enhanced (DUD-E)\cite{DUD-E} and Demanding Evaluation Kits for Objective \textit{In silico} Screening (DEKOIS) 2.0\cite{DEKOIS2.0} benchmarks to evaluate the performance of PIGNet2 in situations resembling the real-world virtual screening scenarios.

\subsubsection{Virtual screening benchmark}
\label{sec:3.2.1}

The DUD-E benchmark is a widely used one for evaluating the virtual screening performance of PLI prediction models.
Compared to CASF-2016, this benchmark is more similar to a real situation as it contains only a few hundred actives for each target in a total of 102 targets, while each active has 50 decoys with similar physico-chemical properties.
While some studies have criticized the DUD-E benchmark for its lack of generalizability,\cite{Challenge2,hiddenbias} most of these investigations are based on the results of both training and inference conducted using the DUD-E benchmark.
We expect that generalizability issues may be attenuated in our case because our model, PIGNet2, was not trained on the DUD-E data.
We additionally adopted DEKOIS2.0 which comprises 81 different targets.
Similar to the DUD-E benchmark, DEKOIS2.0 contains dozens of active compounds and thousands of decoys for each target.

To evaluate the screening performance of PIGNet2 and to conduct a comparative assessment with other deep learning-based PLI prediction models, we selected the top $\alpha$ \% enrichment factor as our primary benchmark metric, consistent with the CASF-2016 screening benchmark.
Additionally, to conduct an ablation study on data augmentation in the context of screening performance, we utilized the Kullback--Leibler (KL) divergence, $D_{KL}$.\cite{KLdiv}
The KL divergence measures the extent to which two distributions diverge.
Here, we used this metric to measure the deviation between the predicted binding affinity distributions of actives and decoys, as the larger the deviation, the easier it is to distinguish between the two. 
Given the predicted value distribution for actives, $D_\mathrm{active}$, and the predicted value distribution for decoys, $D_\mathrm{decoy}$, the KL divergence is represented by the following equation:
\begin{equation}
\label{eq:KL_divergence}
D_{KL} \left(D_\mathrm{active} \Vert D_\mathrm{decoy}\right)= \sum_{x\in\chi} D_\mathrm{active}(x) \log\left(\frac {D_\mathrm{decoy}(x)}{D_\mathrm{active}(x)} \right).
\end{equation}
The KL divergence always has a positive value, and the higher its value, the greater the deviation between the $D_\mathrm{active}$ and $D_\mathrm{decoy}$ distributions.

\subsubsection{Baseline models}
\label{sec:3.2.2}
We adopted three deep learning-based PLI prediction models for the comparative analysis.
OnionNet-SFCT is a model that predicts correction terms for the scoring function of docking programs, as previously mentioned in Section \ref{sec:3.1.2}.
For the comparison, we chose the model that uses a scoring function correction term for Autodock Vina.\cite{AutodockVina}
GenScore predicting the distance likelihood was one of the best PLI prediction models in the screening performance.
GNINA is a 3D CNN-based model that is trained using various data augmentation strategies and tasks like PIGNet2.
GNINA offers several PLI prediction models that mainly target virtual screening.
Here, we chose the Dense (Affinity) model, namely Dense, which showed state-of-the-art results among exact binding affinity prediction models in a previous research.\cite{vsgnina}

\begin{figure*}[t]
 \centering
 \includegraphics[width=\textwidth]{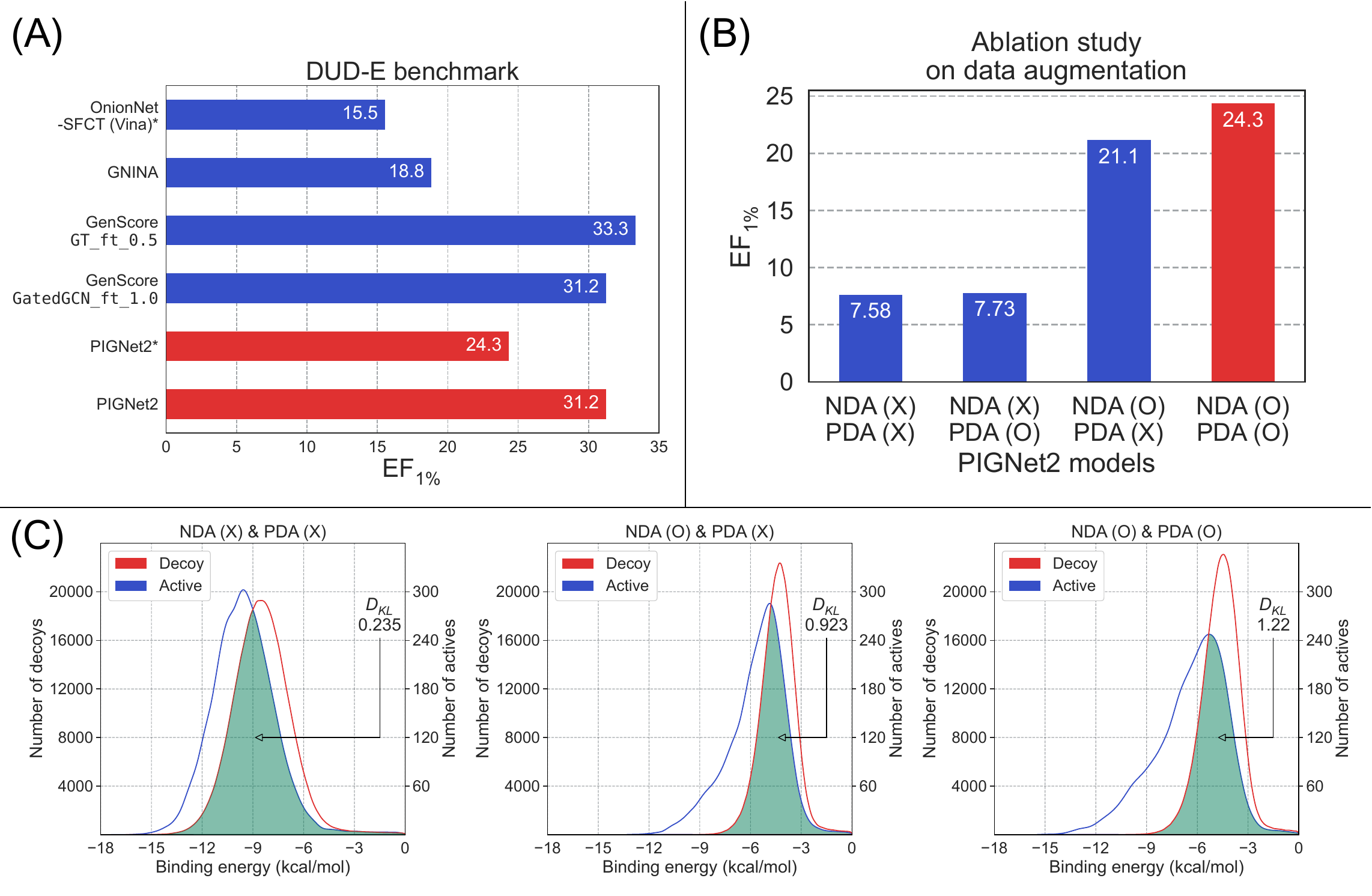}
 \caption{
 Results for the DUD-E benchmark.
 $\mathrm{EF}_{1\%}$ is the top 1\% enrichment factor.
 (A) Comparison between baseline models and PIGNet2 in terms of the $\mathrm{EF}_{1\%}$.
 The models with the asterisk (*), OnionNet-SFCT (Vina) and PIGNet2, used a single conformation generated by docking, while the others used multiple conformation.
 (B) Ablation study about data augmentation strategies for PIGNet2 on the enrichment factor of the DUD-E benchmark.
 We conducted all experiments in the ablation study with the same dataset used in the result of PIGNet2* in (A), which employed only the best pose prepared by Smina for each complex in the DUD-E benchmark.
 Each labels on x-axis means the data augmentation strategies used for training PIGNet2.
 For example, NDA (O) means the model trained with NDA and PDA (X) means the model trained without PDA.
 The red bar shows PIGNet2 model with the best performance.
 (C) Ablation study about data augmentation strategies for PIGNet2 on predicted distributions of actives and decoys of the DUD-E benchmark.
 Three models are compared: the model trained without any data augmentation strategies (left), the model trained with NDA alone (middle), and the model trained with both NDA and PDA (right).
 In each sub-figure, the green colored region is an overlap between the distributions of actives and decoys.
 }
 \label{fig:dude_benchmark}
\end{figure*}

\subsubsection{Performance of PIGNet2}
\label{sec:3.2.3}
Figure \ref{fig:dude_benchmark}(A) shows the average DUD-E top 1\% enrichment factor, $\mathrm{EF}_{1\%}$, for PIGNet2 and the baseline models.
For the evaluation of the screening performance of PIGNet2, we rescored the docking-generated structures of actives and decoys for a given target provided by GenScore.\cite{GenScore}
The top 10 structures originally generated and scored using Glide SP underwent further minimization with Smina.
PIGNet2 showed much higher performance than OnionNet-SFCT~(Vina) and GNINA, which is close to the state-of-the-art results of GenScore.
It is worth noting that these results could be influenced by potential biases arising from different numbers of sampled structures with different sampling algorithms for each method.
For instance, the result for OnionNet-SFCT~(Vina) was computed based on the best binding pose generated by Autodock Vina, whereas GNINA used up to nine structures generated by its own algorithm, and GenScore and PIGNet2 used 10 structures as described above. 

To mitigate such possible biases, we further investigated the performance of PIGNet2 by using only the best pose generated by Smina and compared the results with those of OnionNet-SFCT~(Vina).
The resulting EF$_{1\%}$, 24.3, was depicted as PIGNet2* in Figure~\ref{fig:dude_benchmark}(A), which was significantly dropped from the result with the top 10 structures (31.2) but is still far better than OnionNet-SFCT~(Vina).

Table \ref{tab:dekois} shows the average top 0.5\%, 1.0\%, and 5.0\% enrichment factors for the DEKOIS2.0 benchmark.
For a fair evaluation, we utilized structures provided from \citet{GenScore}, since all values of the baseline models were taken from \citet{GenScore}.
For each complex, the top ten structures were generated by Glide SP\cite{GlideSP} in the original dataset, but we further minimized them using Smina for PIGNet2.
PIGNet2 outperformed Glide SP, thereby validating its superior capability in the virtual screening task.
For $\mathrm{EF}_{0.5\%}$, PIGNet2 demonstrated competitive performance to GenScore (\texttt{GT\_ft\_0.5}), an optimal one for screening among its various versions, while surpassing GenScore (\texttt{GatedGCN\_ft\_1.0}), an optimal model for scoring. As for $\mathrm{EF}_{1\%}$ and $\mathrm{EF}_{5\%}$, PIGNet2 showed slightly better performance than both versions of GenScore. 

\begin{table}[h!]
\caption{
Results of the DEKOIS2.0 benchmark. EF$_{0.5\%}$, EF$_{1.0\%}$, and EF$_{5.0\%}$ are average top 0.5\%, 1.0\%, and 5.0\% enrichment factor, respectively.
For PIGNet2, we report the results of the ensemble of 4 randomly initialized models trained with both positive data augmentation and negative data augmentation.
The results of the other models were taken from \citet{GenScore}
}
\centering
\begin{tabular}{lccc}
\toprule
Model & EF$_{0.5\%}$ & EF$_{1.0\%}$ & EF$_{5.0\%}$ \\
\midrule
Glide SP & 14.6 & 12.5 & 6.30 \\
GenScore (\texttt{GT\_ft\_0.5}) & 20.2 & 17.9 & 8.25 \\
GenScore (\texttt{GatedGCN\_ft\_1.0}) & 18.6 & 17.0 & 7.93 \\
\midrule
PIGNet2 & 20.0 & 18.6 & 9.71 \\
\bottomrule
\end{tabular}
\label{tab:dekois}
\end{table}

\subsubsection{Ablation study on data augmentation strategies}
\label{sec:3.2.4}
To elucidate the impact of our proposed data augmentation strategies on screening performance, we conducted ablation studies based on $\mathrm{EF}_{1\%}$ results of models trained with and without NDA and PDA in the DUD-E benchmark.
This is detailed in Figure \ref{fig:dude_benchmark}(B).
It is noteworthy that we employed the same dataset used to compare the EF$_{1\%}$ of PIGNet2 with OnionNet-SFCT (Vina), which used only the best structure for each ligand generated by Smina for all ablation studies since the provided data from GenScore does not fully contain all of the original data in the DUD-E benchmark.

First, we observed that a model solely trained on crystal structures without both PDA and NDA exhibited no increase in $\mathrm{EF}_{1\%}$ after incorporating PDA.
Since PDA helps the model to observe near-native conformations of binders, rather than discriminating diverse decoys as NDA does, this result could be expected.
However, adding PDA to a model already improved by NDA further improved its performance compared to a model trained with NDA alone.
The observation that the gain in $\mathrm{EF}_{1\%}$ from the introduction of PDA is more prominent in the NDA-only model than in the model without both PDA and NDA may suggest that PDA can differentiate more of the active and decoy structures that are rendered indistinguishable in the NDA-only model than in a model trained without any data augmentation.
This suggests that PDA can effectively regularize the potential biases introduced by NDA, enhancing the model to discriminate between actives and decoys.
Ultimately, when we compared the $\mathrm{EF}_{1\%}$ of the model trained without any data augmentation and the model trained with all data augmentation methods, the result of the latter was more than twice higher than that of the former.

Next, we plotted the distribution of predicted binding affinity for actives and decoys in the DUD-E benchmark, as shown in Figure \ref{fig:dude_benchmark}(C), to analyze how the use of each data augmentation leads to an improvement in $\mathrm{EF}_{1\%}$.
The model trained without any data augmentation is on the left, the one trained with only NDA is in the middle, and the one trained with both NDA and PDA is on the right.
Without data augmentation, the majority of the active and decoy distributions overlap.
However, the overlap decreases when data augmentation is employed, implying that the active and decoy distributions diverge more significantly.

Interestingly, the distribution of predictions from the model with NDA shifted to the right compared to the model without any data augmentation.
This could be associated with the hinge loss used in most models that employ NDA, which results in lower binding affinity predictions for some active molecules, consequently weakening both the screening and scoring performance.
When PDA was additionally employed in the training process of the model trained with only NDA, there was still some shift in the active distribution compared to the model without any data augmentation.
However, this shift was less than that of the model trained with only NDA, and the overall distributions of actives and decoys are more clearly distinguished.

Moreover, the number of actives in high binding affinity regions relative to decoys, which is the left tail of the blue-colored active distribution, increased progressively from left to right in Figure \ref{fig:dude_benchmark}(C).
This result directly accounts for the increase in $\mathrm{EF}_{1\%}$, as $\mathrm{EF}_{1\%}$ measures the proportion of active molecules in the top 1\% of predictions.
Therefore, the relative number of actives in high binding affinity regions increased the $\mathrm{EF}_{1\%}$ value.

\begin{table*}[t]
\caption{
Results on the derivative benchmark 2015, where $R$ means Pearson correlation coefficient.
The results of the models with the asterisk (*) are from \citet{DerivativeBenchmark}, while the others were calculated in this work.
For PIGNet2, we report the results of the ensemble of 4 randomly initialized models trained with both positive data augmentation and negative data augmentation
}
\resizebox{\textwidth}{!}
{
\centering
\begin{tabular}{lccccccccc}
\toprule

\multirow{2.5}{*}{Model} &
\multirow{2.5}{*}{\shortstack[c]{\strut Performance\\\strut Average $R$}} &
\multicolumn{8}{c}{Systems} \\
\cmidrule{3-10} 

 &&
  \multicolumn{1}{c}{BACE} &
  \multicolumn{1}{c}{CDK2} &
  \multicolumn{1}{c}{JNK1} &
  \multicolumn{1}{c}{MCL1} &
  \multicolumn{1}{c}{p38} &
  \multicolumn{1}{c}{PTP1B} &
  \multicolumn{1}{c}{Thrombin} &
  \multicolumn{1}{c}{TYK2} \\
\midrule
MM-GB/SA\cite{DerivativeBenchmark}* & 0.40 & -0.40 & -0.53 & 0.65  & 0.42 & 0.66 & 0.67 & 0.93 & 0.79 \\
Glide SP\cite{DerivativeBenchmark}* & 0.29 & 0.00  & -0.56 & 0.24  & 0.59 & 0.14 & 0.55 & 0.53 & 0.79 \\
Smina\cite{Smina} & 0.25 & -0.48 & 0.10  & -0.060 & 0.24 & 0.52 & 0.70 & 0.72 & 0.24 \\
\midrule
OnionNet-SFCT (Vina)\cite{OnionNet-SFCT} & 0.023 & -0.48 & -0.68 & -0.59 & 0.29 & 0.50 & 0.66 & 0.71 & -0.23 \\
Sfcnn\cite{Sfcnn} & 0.084 & -0.24 & 0.044 & -0.65 & 0.12 & 0.58 & 0.58 & 0.041 & 0.20 \\
GenScore (\texttt{GT\_ft\_0.5})\cite{GenScore}* & 0.57 & 0.45 & 0.63 & 0.63 & 0.54 & 0.61 & 0.52 & 0.92 & 0.25 \\
GenScore (\texttt{GatedGCN\_ft\_1.0})\cite{GenScore}* & 0.57 & 0.35 & 0.62 & 0.71 & 0.47 & 0.65 & 0.65 & 0.88 & 0.22 \\
\midrule
PIGNet2 & 0.64 & 0.42 & 0.77 & 0.36 & 0.78 & 0.60 & 0.76 & 0.83 & 0.61 \\
\bottomrule
\end{tabular}
}
\label{tab:Derivative_benchmark}
\end{table*}

To complement the analysis above, we evaluated the separation between the predicted binding affinity distributions of active and decoy instances in terms of $D_{KL}$ as a result of data augmentation.
Despite not being a direct assurance of high $\mathrm{EF}_{1\%}$, an greater $D_{KL}$ can be suitably considered as a strong indicator positively associated with a high $\mathrm{EF}_{1\%}$.
The rationale behind this connection lies in the fact that a larger $D_{KL}$ signifies a greater distinction between active and decoy distributions, thus enhancing the likelihood of identifying more actives in regions characterized by higher binding affinities compared to decoys. 
As depicted in Figure \ref{fig:dude_benchmark}(C), the $D_{KL}$ of the model without any data augmentation, the NDA-only model, and the model with both NDA and PDA are 0.235, 0.923, and 1.22, respectively.
In summary, we can conclude that adding NDA and PDA can better separate the predicted binding affinity distributions of active and decoy.

\subsection{Performance on ranking structurally similar compounds}
\label{sec:3.3}
Selecting molecules with higher binding affinity to a target among plenty of similar derivatives is an important task during hit-to-lead and lead optimization.
For this purpose, one can rank the relative binding affinities of similar derivative molecules.
However, this remains challenging due to issues such as activity cliffs, where small changes in the molecule can result in significant changes in activity.
To further demonstrate the exceptional performance of PIGNet2 in predicting binding affinity, especially when compared to other PLI prediction models, we evaluated the predictive performance of the others on derivative benchmarks.

\subsubsection{Derivative benchmark}
\label{sec:3.3.1}
We considered two sets of derivative benchmarks reported in 2015\cite{DerivativeBenchmark} and 2020,\cite{merck} respectively. The derivative benchmark 2015 is composed of 199 derivatives and their corresponding experimental binding energies for eight target systems. 
The derivative benchmark 2020 is similar to the derivative benchmark 2015 and comprises a total of 264 active ligands for eight targets.
We leveraged this data to evaluate the ability to predict the relative binding affinity among similar derivatives for a given target, which is assessed by calculating the Pearson correlation coefficient $R$ between predicted and experimental values.

\subsubsection{Baseline models}
\label{sec:3.3.2}
\citet{DerivativeBenchmark} compared the binding affinity prediction performance of several physics-based methods: free energy perturbation (FEP), molecular mechanics with generalized Born and surface area solvation (MM-GB/SA), and Glide SP.
Here, we aimed to evaluate methods with comparable computational costs, excluding FEP due to its high computational cost despite its high accuracy.
The remaining MM-GB/SA and Glide SP results are computed from a single snapshot of a given complex structure.
In particular, Glide SP is a widely-used traditional scoring function.
We also introduce another traditional scoring function, Smina,\cite{Smina} which is used for the minimization of our initial structures.

Moreover, to compare with the other deep learning methods, we only considered models for which the provided code can be applied or for which benchmark results were already available. As a result, we selected Sfcnn,\cite{Sfcnn} OnionNet-SFCT (Vina),\cite{OnionNet-SFCT} and GenScore.\cite{GenScore}
These models adopted different approaches as discussed in the CASF-2016 benchmark.
In summary, MM-GB/SA, Glide SP, and Smina can be categorized as traditional scoring functions, while GenScore, OnionNet-SFCT (Vina), and Sfcnn are deep learning-based PLI prediction models.

For the derivative benchmark 2015, the result of OnionNet-SFCT (Vina) is obtained by rescoring the docked structures from Autodock Vina.
For both GenScore models (\texttt{GT\_ft\_0.5} and \texttt{GatedGCN\_ft\_1.0}), we used the provided structures as-is.
The results of all other models were obtained using the structures minimized by Smina.
For the derivative benchmark 2020, the results of MM-GB/SA, Glide SP, and Vina are from the \citet{GenScore}
The provided structures were generated by the Flexible Ligand Alignment tool or Glide core-constrained docking using the respective reference structure.
We again minimized the given structures using Smina when evaluating PIGNet2.

\begin{table*}[t]
\caption{
Results on the derivative benchmark 2020, where $R$ means Pearson correlation coefficient.
Except for PIGNet2, all the results are originated from \citet{GenScore}
For PIGNet2, we report the results of the ensemble of 4 randomly initialized models trained with both positive data augmentation and negative data augmentation
}
\resizebox{\textwidth}{!}
{
\centering
\begin{tabular}{lccccccccc}
\toprule

\multirow{2.5}{*}{Model} &
\multirow{2.5}{*}{\shortstack[c]{\strut Performance\\\strut Average $R$}} &
\multicolumn{8}{c}{Systems} \\
\cmidrule{3-10} 

 &&
  \multicolumn{1}{c}{HIF2$\alpha$} &
  \multicolumn{1}{c}{PFKFB3} &
  \multicolumn{1}{c}{Eg5} &
  \multicolumn{1}{c}{CDK8} &
  \multicolumn{1}{c}{SHP2} &
  \multicolumn{1}{c}{SYK} &
  \multicolumn{1}{c}{c-Met} &
  \multicolumn{1}{c}{TNKS2} \\
\midrule
MM-GB/SA\cite{GenScore} & 0.35 & 0.28 & 0.55 & -0.002 & 0.65 & 0.59 & 0.11 & 0.50 & 0.16 \\
Glide SP\cite{GenScore} & 0.30 & 0.45  & 0.48 & -0.11 & 0.35 & 0.54 & -0.006 & 0.38 & 0.32 \\
Vina\cite{GenScore} & 0.34 & 0.49 & 0.55  & -0.52 & 0.85  & 0.57 & 0.52 & -0.26 & 0.54 \\
\midrule
GenScore (\texttt{GT\_ft\_0.5})\cite{GenScore} & 0.47 & 0.36 & 0.45 & 0.21 & 0.67 & 0.61 & 0.23 & 0.69 & 0.54 \\
GenScore (\texttt{GatedGCN\_ft\_1.0})\cite{GenScore} & 0.52 & 0.52 & 0.58 & 0.21 & 0.71 & 0.61 & 0.21 & 0.73 & 0.59 \\
\midrule
PIGNet2 & 0.43 & 0.45 & 0.29 & -0.09 & 0.37 & 0.72 & 0.50 & 0.57 & 0.64 \\
\bottomrule
\end{tabular}
}
\label{tab:Merck_benchmark}
\end{table*}

\begin{table*}[t!]
\caption{
Ablation studies for PIGNet2 on the derivative benchmark 2015, where $R$ means Pearson correlation coefficient.
For each model, we report the results of the ensemble of 4 randomly initialized models.
The best performance is shown in bold
}
\resizebox{\textwidth}{!}
{
\centering
\begin{tabular}{lccccccccccc}
\toprule

\multirow{2.5}{*}{Model} &
\multicolumn{2}{c}{Data augmentation} &
\multirow{2.5}{*}{\shortstack{\strut Performance\\\strut Average $R$}} &
\multicolumn{8}{c}{Systems} \\

\cmidrule{2-3} \cmidrule{5-12}

&
\multicolumn{1}{c}{Negative} &
\multicolumn{1}{c}{Positive} &
&
\multicolumn{1}{c}{BACE} &
\multicolumn{1}{c}{CDK2} &
\multicolumn{1}{c}{JNK1} &
\multicolumn{1}{c}{MCL1} &
\multicolumn{1}{c}{p38} &
\multicolumn{1}{c}{PTP1B} &
\multicolumn{1}{c}{Thrombin} &
\multicolumn{1}{c}{TYK2} \\
\midrule
PIGNet2 & X & X & 0.50 & -0.16 & 0.36 & 0.21 & 0.71 & 0.67 & 0.64 & 0.82 & 0.74 \\
PIGNet2 & X & O & 0.54 & 0.23 & 0.61 & 0.33 & 0.69 & 0.64 & 0.76 & 0.66 & 0.37 \\
PIGNet2 & O & X & 0.39 & 0.085 & -0.29 & 0.25 & 0.75 & 0.45 & 0.32 & 0.82 & 0.71 \\
PIGNet2 & O & O & \textbf{0.64} & 0.42 & 0.77 & 0.36 & 0.78 & 0.60 & 0.76 & 0.83 & 0.61 \\
\bottomrule
\end{tabular}
}
\label{tab:Derivative_benchmark_ablation}
\end{table*}

\subsubsection{Performance of PIGNet2}
\label{sec:3.3.3}
The overall performance of all models on the derivative benchmark 2015 is presented in Table \ref{tab:Derivative_benchmark}.
PIGNet2 outperformed all the other models.
Glide SP and Smina showed anti-correlated results for specific systems such as BACE and CDK2.
Given that the former was developed for virtual screening, its limited performance in binding affinity scoring was expected.
Remarkably, PIGNet2 showed better performance than MM-GB/SA, which is based on molecular dynamics and hence expected to be more accurate than the docking methods.
Even for the BACE and CDK2 systems where MM-GB/SA displayed anti-correlated tendencies, PIGNet2 shows a positive correlation.

Since the derivative benchmark requires accurate prediction of the binding affinity of the structurally similar derivatives for the same target protein, it is a much more challenging task than scoring and ranking in the CASF-2016 benchmark, where protein-ligand complexes vary significantly.
OnionNet-SFCT (Vina), developed based on $\Delta$ learning focusing on virtual screening, showed anti-correlated results in almost all systems, albeit with its impressive docking power.
Surprisingly, Sfcnn, designed to score binding affinities accurately and has the exact binding affinity as its prediction target, performed only marginally better than OnionNet-SFCT (Vina) and poorer than traditional scoring functions.
This unsatisfactory performance may be because Sfcnn, as a 3D CNN-based model, was trained exclusively on crystal structures and thus struggled to accurately score structures optimized by Smina.
Both GenScore models showed much better performances with no anti-correlation for all the targets as expected, which have a slightly lower average $R$ value (0.57) than PIGNet2 (0.64).

Furthermore, the performance of PIGNet2 in the derivative benchmark 2015, especially in terms of average $R$, is surprisingly close to that of PBCNet\cite{PBCNet} (0.65), a model exclusively designed for predicting the relative binding affinity of two given derivatives.
This observation underscores the competitive position of PIGNet2 within the derivative benchmark, particularly when compared to models designed for predicting relative binding affinities.
Remarkably, the superior performance of PIGNet2 persists even when predicting the binding affinity for each complex apart from the relative value, thereby demonstrating the effective binding affinity prediction performance of PIGNet2, which can also make it applicable to other tasks such as virtual screening.

Considering the encouraging results of PIGNet2 in the derivative benchmark 2015, we expected similar success in the derivative benchmark 2020.
As detailed in Table~\ref{tab:Merck_benchmark}, our model exhibited a remarkable ability when compared to traditional physics-based methods.
However, the average $R$ value of PIGNet2 (0.43) showed a marginally lower performance compared to the best of GenScore (0.52), while showing compatibility with the PBCNet (0.47).
In the evaluation of eight targets, PIGNet2 showed positive correlations for seven systems, while a negative correlation was observed for the Eg5 system.
Interestingly, for this particular system, a more pronounced negative correlation was observed in other physics-based methods, suggesting that physics-based approaches may have a peculiar difficulty for this system.

\begin{figure}[h!]
 \centering
 \includegraphics[width=0.40\textwidth]{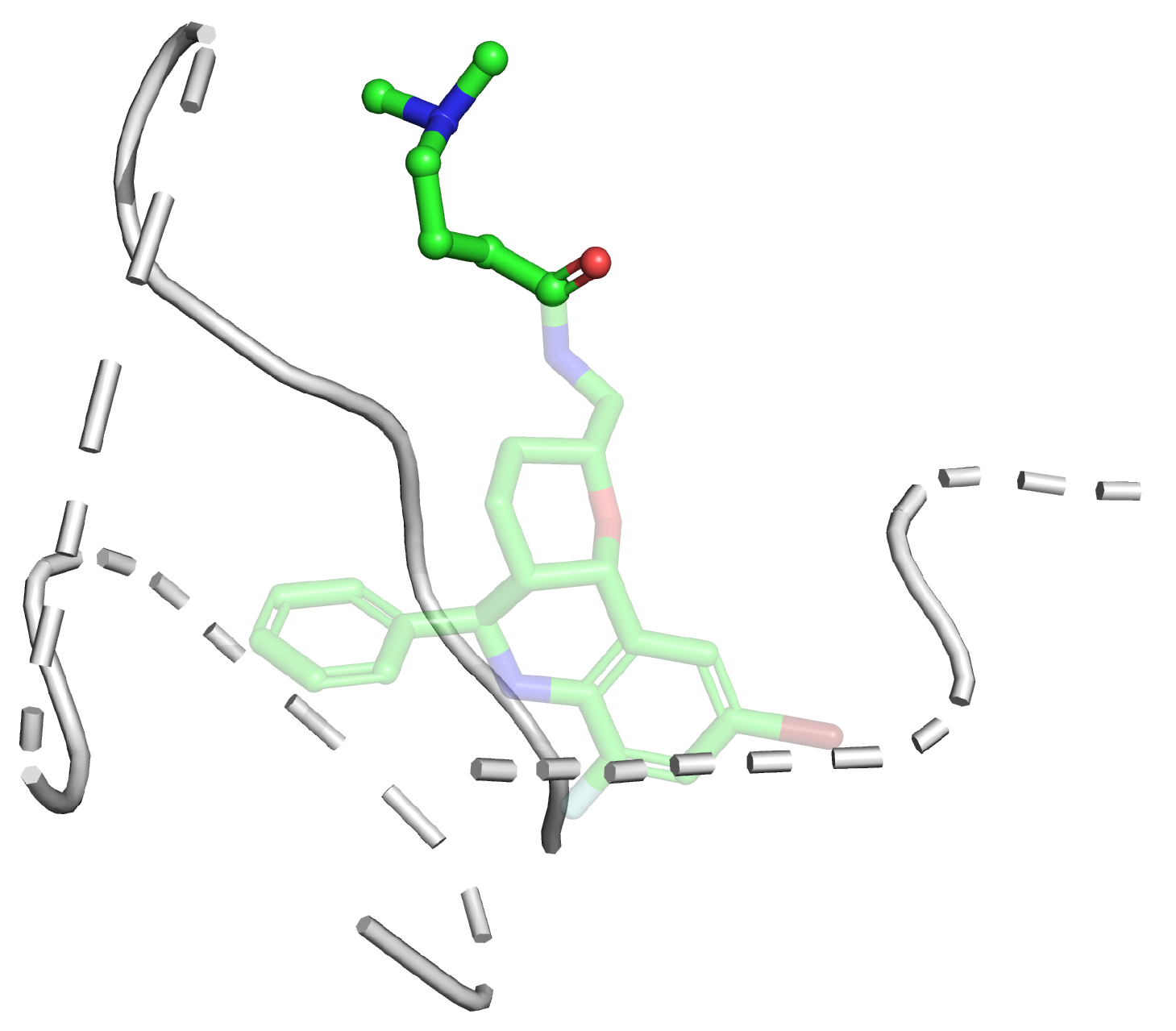}
 \caption{
 Illustration of an Eg5 complex with the ligand CHEMBL1085692, in which the predicted binding affinity of PIGNet2 is lower than the others, resulting in a negative contribution to the correlation between predicted and experimental binding affinities.
 In the illustration, the transparent part is the base molecule, while the other part depicted with a ball-and-stick model is a flexible chain attached.
 The molecular geometries are plotted with PyMol.\cite{PyMOL}
 }
 \label{fig:eg5_case_study}
\end{figure}

By virtue of the explainability of physics-based approaches, we performed further analysis on why the anti-correlation occurred in the Eg5 system.
A previous report on the derivative benchmark 2020\cite{merck} indicated that adding a flexible chain to the base molecule generally leads to better binding affinity.
In contrast to this report, PIGNet2 predicted lower binding affinities to molecules with the flexible chain, causing a negative correlation to be observed.
As Figure~\ref{fig:eg5_case_study} illustrates, the flexible chain prefers to head out to the water rather than fit into the protein pocket, leaving the binding molecule more exposed.
Because empirical physics terms used in PIGNet2 consider solvation effects only implicitly, it is expected that the model may less accurately predict the binding affinity of molecules that are more exposed to water, as is observed from the Eg5 system.
This analysis highlights the great advantage of physics-based models with high explanatory power, which then provides directions for further improvement.

\subsubsection{Ablation study on data augmentation strategies}
\label{sec:3.3.4}
To understand the impact of PDA on scoring performance, we conducted an ablation study for PIGNet2 on the derivative benchmark 2015.
Table~\ref{tab:Derivative_benchmark_ablation} provides a comprehensive comparison of the influence of NDA and PDA on the scoring performance of PIGNet2 across multiple datasets, with the effect of PDA particularly highlighted in terms of the average Pearson correlation coefficient ($R$).
An initial comparison between the model trained without any data augmentation and the one trained solely with NDA revealed a slight decrease in the overall average $R$.
Notably, we can observe an inversion of the correlation for CDK2 due to the use of NDA, which negatively affects the scoring performance, while changes in other systems remained relatively negligible.

However, when PDA was incorporated alongside NDA, the model regained a high correlation for CDK2.
Moreover, the model trained exclusively with PDA alone outperformed the model trained without any data augmentation in terms of scoring performance.
These results suggest a significant impact of PDA in mitigating potential biases that could occur when training with only crystal structures and NDA alone, thereby significantly improving the scoring performance of PIGNet2.

\begin{figure}[h!]
 \centering
 \includegraphics[width=0.48\textwidth]{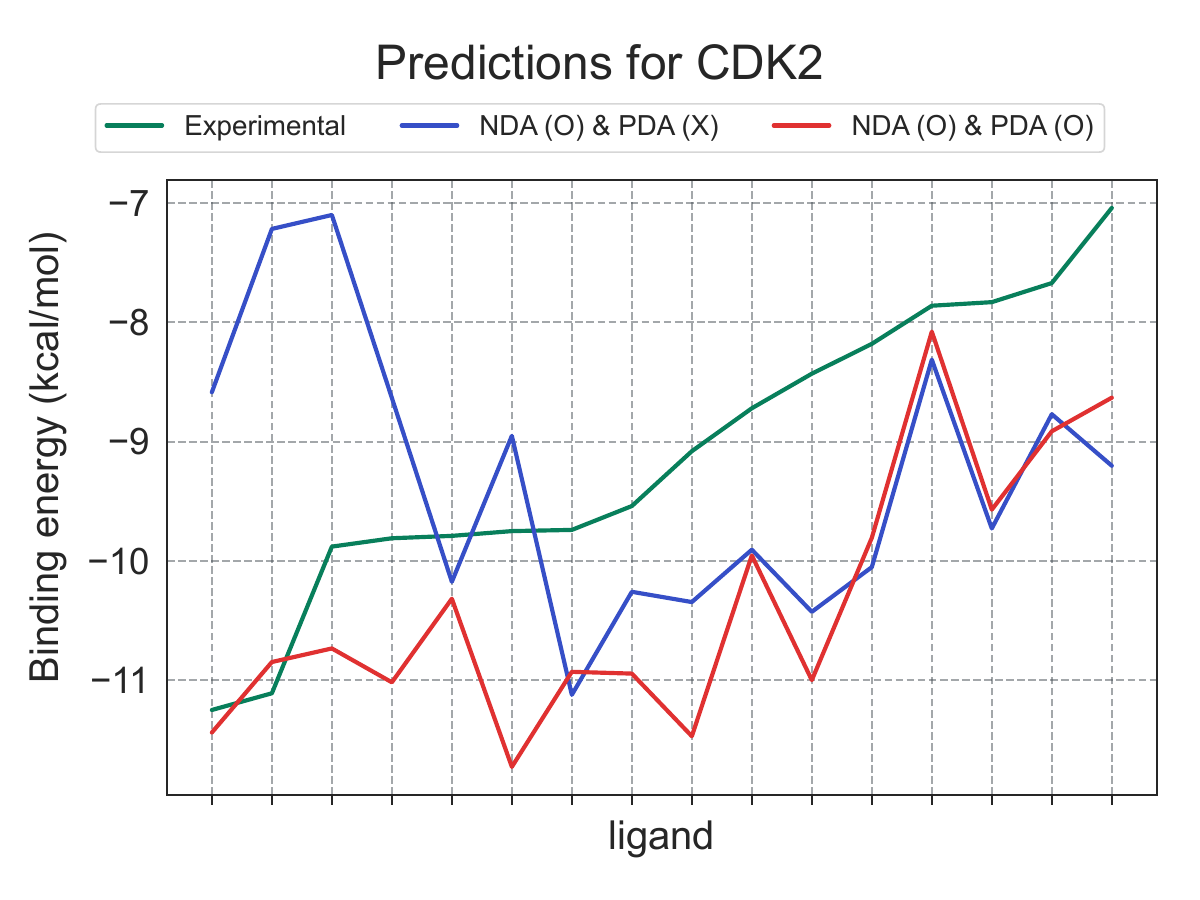}
 \caption{
 A case study for a CDK2 target system for ablation study about PDA on the derivative benchmark.
 The CDK2 system has 16 derivatives in total, and the illustrated result is sorted based on the experimental binding energy for all derivatives.
 Specifically, this figure additionally illustrates the prediction results of two models: the model trained with NDA alone and the model trained with both NDA and PDA.
 }
 \label{fig:derivative_ablation_cdk2}
\end{figure}

The incorporation of PDA manifests itself as either maintaining or positively affecting most systems except TYK2, with some systems showing particularly profound improvements.
In this context, we have visualized the affinity prediction results for CDK2, the target for which the introduction of PDA led to the most significant performance improvements, in Figure \ref{fig:derivative_ablation_cdk2}.
In particular, the figure highlights a drastic reversal of the correlation trend from -0.29 to 0.77 upon incorporating PDA into a model initially trained on NDA alone.
This transformation may illustrate the immense potential of PDA in enabling a deep learning-based PLI prediction model to improve scoring performance and overcome limitations encountered when using NDA alone.

\section{Conclusions}
In this study, we present PIGNet2, a versatile deep learning-based protein-ligand interaction (PLI) prediction model that enhances its generalization ability with appropriate physics-based inductive bias and, in addition to negative data augmentation (NDA), a novel data augmentation strategy called positive data augmentation (PDA).
Unlike NDA, PDA generates near-native structures treated as equivalent to crystal structures during training.
PIGNet2 incorporates both NDA and PDA, enabling accurate binding affinity prediction for near-native structures and effective discrimination between active and decoy molecules.
Remarkably, PIGNet2 outperformed task-specific deep learning models and traditional physics-based methods in all benchmarks and is on par with the state-of-the-art performance reported recently. Furthermore, it has the distinctive advantage of predicting exact binding affinities using intuitively explainable physics, allowing for direct comparison to experimental results and providing directions for further improvement.
This result solidifies the potential of PIGNet2 as a versatile deep learning-based PLI prediction model suitable for both scoring and screening tasks in drug discovery.

Despite its high potential, the present method has room for further improvement.
First, our generation procedure of binding structures for data augmentation would be biased toward the scoring function of the Smina docking software.
This bias results in challenges when dealing with structures generated by other methods.
Second, the use of NDA with the hinge loss could lead to lower predicted binding affinities for actives, as observed in the DUD-E benchmark ablation study.
Third, better solvation effects should be considered to improve the prediction accuracy for molecules that are more exposed to water as observed in the benchmark result of the eg5 system. Fourth, exploring better physics terms for energy evaluation is expected to enhance the performance significantly, as the present physics terms adopted from conventional docking methods impose strong constraints in the training process and thus limit the power of deep learning.
Future studies may address these issues to enhance the reliability of PLI prediction models in the drug discovery process.

\section*{Data availability}
The code and trained models are available at github: \url{https://github.com/ACE-KAIST/PIGNet2}.
Also, data is available at \url{https://doi.org/10.5281/zenodo.8091220}.

\section*{Author Contributions}
Conceptualization: S.M. and W.Y.K.; methodology: S.M. and J. L.; software, investigation, and formal analysis: S.M., S.-Y.H.; writing – original draft S.M.; writing – review \& editing: S.M., S.-Y.H., J.L., and W.Y.K.; supervision: W.Y.K.

\section*{Conflicts of interest}
There are no conflicts to declare.

\section*{Acknowledgements}
This work was supported by the National Research Foundation of Korea (NRF) grant funded by the Korea government (MSIT) (RS-2023-00257479).



\balance


\bibliography{rsc} 
\bibliographystyle{rsc} 

\end{document}